\documentstyle[graphics,amsmath,epsfig]{europhys}
\def\stars{\bigskip\centerline{***}\medskip}

\newif\ifboo \boofalse

\newcommand{\vmu}[1]{\widetilde{\mathbf{#1}}}

\renewcommand{\tilde}[1]{\widetilde{#1}}

\begin{document}
\euro{00}{0}{000-000}{2000}
\Date{00 00 2001}
\shorttitle{A.G. MOREIRA ET AL COUNTERIONS AT NANO-STRUCTURED SUBSTRATES}

\title{Counterions at charge-modulated substrates}

\author{
Andr{\'e} G. Moreira$^{1,2}$ and Roland R. Netz$^{1,3}$}

\institute{$^{1}$Max-Planck-Institut f{\"u}r Kolloid- 
  und Grenzfl{\"a}chenforschung, 14424 Potsdam, Germany \\
  $^{2}$Materials Research Laboratory, UCSB, Santa Barbara, CA 93106, USA\\
  $^{3}$Sektion Physik, LMU, Theresienstr. 37, 80333 M\"unchen, Germany}

\rec {00 00 2001}{}

\pacs{
  \Pacs{82}{70.-y}{Disperse systems; complex fluids}
  \Pacs{61}{20.Qg}{Structure of associated liquids: electrolytes, molten salts, etc.}
  \Pacs{82}{45.+z}{Electrochemistry}
}

\maketitle

\begin{abstract}
We consider counterions in the presence of a single
planar surface with a spatially inhomogeneous
charge distribution using Monte-Carlo simulations and strong-coupling theory. 
For high surface charges, multivalent 
counterions, or pronounced substrate charge modulation the counterions are laterally 
correlated with the surface charges and their density profile deviates
strongly from the limit of a smeared-out substrate charge distribution, 
in particular exhibiting a much increased laterally averaged density at the surface. 
\end{abstract}

The recently revived interest in charged soft-condensed matter systems reflects
that there are still many open questions, 
despite the enormous amount of work available in this field. 
One example of high experimental relevance is
 the discrete nature of  charged surface groups,
or, more generally, the inhomogeneity of substrate charge distributions,
and how it affects various thermodynamic properties in an aqueous environment
such as forces between
charged particles or the counterion distribution.
The importance of the discreteness of charged surface groups has been experimentally 
established in colloidal flocculation\cite{kihira:92} and 
deposition studies\cite{litton:94} and was recently reviewed\cite{walz:98}.
On a much larger length scale, chemically micropatterned substrates with charged 
and neutral patches can be used for controlled colloidal deposition\cite{aizenberg-al:2000}
and DNA immobilization\cite{Gaub}.
Theoretically, charge-modulated surfaces have been studied by various mean-field
approximations\cite{richmond:75,chan:80,kostoglou:92,peitzsch-al:95,Shklovskii}, 
liquid state theory\cite{kjellander-marcelja:88,pinc},
as well as computer simulations\cite{vanmegen:80,note:bo,messina:01}. 
In these studies, the importance of inhomogeneous surface charge distributions 
has been recognized, and for a number of different charge-distribution 
models and parameters the ionic distribution functions as well as forces 
between charged surfaces have been calculated.
In this paper we use a simple model for a charge-modulated surface which 
depends on a single geometric parameter and 
includes as limiting cases both smeared-out and delta-peaked 
charge distributions. We study the distribution of counterions without 
added salt at a 
single charged surface for all different values of the electrostatic
coupling parameter (depending on temperature and counterion valence)
and for different degrees of surface charge modulation, both using
Monte-Carlo (MC) simulation methods and the recently introduced strong-coupling 
(SC) theory\cite{moreira-netz:epl,moreira-netz:prl,roland-sc}. Our study therefore
encompasses different experimental situations such as mono/multivalent
ions at surfaces with discrete charged chemical groups, as well as charged 
colloids at microscopically charge-modulated surfaces. 

As we will demonstrate, the assumption of smeared-out charges
dramatically breaks down when the charge-modulation at the surface is 
pronounced, but also for moderate substrate-charge modulation when the
 electrostatic coupling is large 
 (i.e. for highly charged surfaces, low temperatures, or multivalent counterions):
in both cases the counterions become highly correlated with the surface charges and
tend to form a two-dimensional, laterally ordered layer close to the surface. 
A convenient measure for the effects of substrate charge modulation is the
counterion contact density, i.e., the laterally averaged counter-ion 
density at the substrate surface, since it is known exactly in the 
smeared-out case and can be easily determined from 
simulations\cite{moreira-netz:epl}.
We find that
the contact density for charge-modulated substrates
can be much larger than for the smeared-out case, which is in agreement 
with recent experimental measurements on highly charged surfactant
monolayers\cite{teppner}.
Quite surprisingly,  substrate charge modulation tends 
to have a more drastic effect on the counter-ion distribution
than fluctuations and correlations, which have been
the subject of numerous recent studies (see \cite{moreira-netz:epl} and 
references therein). 
As our numerical results show, the SC 
theory describes the counterion distributions quantitatively
in the SC limit and in particular in the limit
of pronounced surface-charge modulation. We also consider
 a dielectric-constant jump at the substrate,
as relevant for charged biological and colloidal 
surfaces\cite{kjellander-marcelja:84,bratko-al:86,
attard-mitchell-ninham:88,podgornik-zeks:88}. In contrast to the
mean-field Poisson-Boltzmann (PB) approach, which severely fails even in the
smeared-out case, our SC approach compares well with numerical results
and helps to understand the intricate interplay of dielectric-jump and 
charge-modulation effects.

\begin{figure}[t]
  \begin{center}
    \epsfig{file=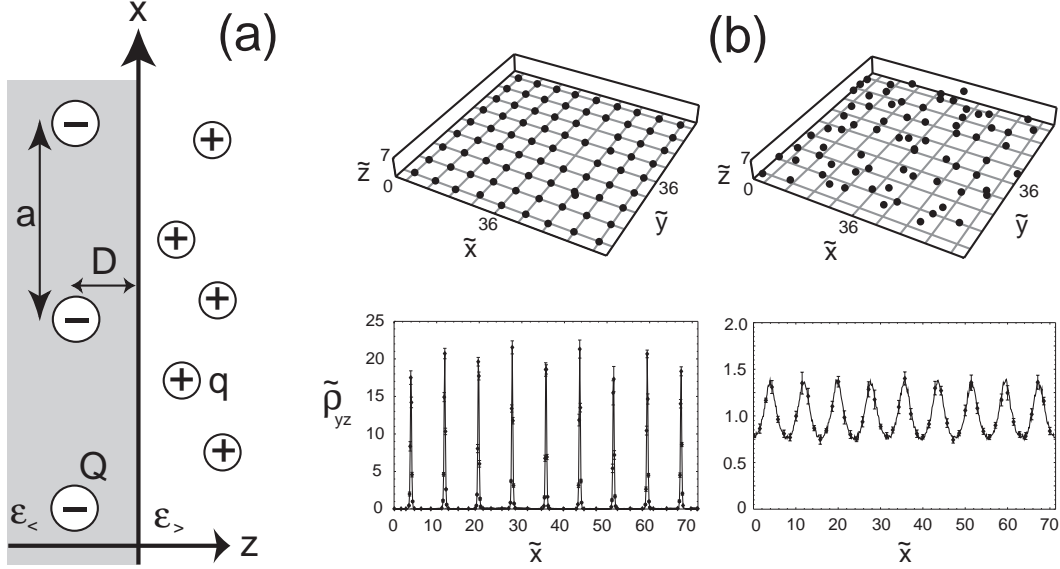, width=140mm}
  \end{center}
  \vspace{-5mm}
  \caption{a) In our model a square lattice of fixed surface charges with 
lattice constant $a$ is located in the $xy$-plane at a distance
$D$ from the counterion containing half space. b)
Simulation snapshots and normalized lateral density profiles 
$\tilde{\rho}_{yz}(\tilde{x})$ 
(averaged in the $yz$--plane) for a coupling constant $\Xi=10$, vanishing 
dielectric jump ($\Delta =0$) and valence symmetry between surface 
 and counterions ($Q=q$) for $D/a = 0.06$ (left) and
$D/a = 0.24$ (right), showing strong and weak lateral counterion
ordering, respectively. 
In the snapshots, the surface ions are located beneath the lattice 
nodes. }
  \vspace{-5mm}
  \label{fig:snapshots}
\end{figure}

The substrate charge distribution is modeled, for simplicity, 
by point charges distributed on a square lattice with lattice constant $a$.
The point-like oppositely charged counterions are confined to the positive
half-space with $z>0$, the minimal distance between surface charges and 
counterions is given by $D$, as shown in Fig.~\ref{fig:snapshots}a.
The dielectric constant of the positive half-space $\varepsilon_>$ is 
 allowed to be different from $\varepsilon_<$.
The system is globally neutral, and only the fixed surface charges
and counterions are present, {\it i.e.}, no salt is added.
In principle, one could interpret $D$ as the sum of the surface-ion and 
counterion hard-core radii.
In our model, however, we use the ratio $D/a$ more generally
to control  the degree of substrate-charge modulation,
 which allows to describe a variety of  
experimental situations and systems by a single parameter.
The limit $D/a \rightarrow \infty$ is equivalent to a smeared-out surface-charge
distribution, while $D/a \rightarrow 0$ corresponds to a delta-peaked 
distribution.

The typical height of the counterion layer in the smeared-out limit
defines the Gouy-Chapman length $\mu = 1 / (2 \pi q \ell_B \sigma_s)$,
where $\sigma_s=Q/a^2$ is the number charge  density at the wall,
$q$ and $Q$ are the valences of counterions and surface ions, 
and $\ell_B = e^2 /4 \pi \varepsilon_> \varepsilon_0 k_B T$ is the 
Bjerrum length (the distance at which two elementary charges interact
with thermal energy, $k_B T$).
In the following, we will rescale all lengths by $\mu$ according 
to $\tilde{r} \equiv r / \mu$. 
For the simple double layer (smeared out
charges at the surface and no dielectric jump) the coupling parameter
$\Xi = q^2 \ell_B / \mu = 2 \pi q^3 \ell_B^2 \sigma_s$ is the only 
parameter in the problem.
In the limit $\Xi \rightarrow 0$ (weakly charged surfaces,
low-valence counterions, or high temperatures) the PB theory is
asymptotically exact, while in the opposite limit $\Xi \rightarrow \infty$ 
(highly charged surfaces,
high-valence counterions, or low temperatures) 
the SC  theory is exact\cite{moreira-netz:epl}. 
The present model  is in addition characterized by
 $D/a$, the dielectric-constant ratio 
$\Delta = (\varepsilon_> - \varepsilon_<) / (\varepsilon_> + \varepsilon_<)$
and the valence ratio $q/Q$  between counterions and surface charges.

The Coulomb interaction in the presence of dielectric discontinuities is given by 
the solution of the Poisson equation with the appropriate 
boundary conditions\cite{jackson:book}.
For a single dielectric jump located at $z=0$ as shown in Fig.1a,
the electrostatic energy reads
\begin{multline}
    \label{energia}
    \frac{{\mathcal H}}{k_B \, T} = \Xi \sum_{i > j}
    \biggl\{ \frac{1}{|\vmu{r}_i - \vmu{r}_j|} + 
    \frac{\Delta}{\sqrt{[\vmu{r}_i - 
        \vmu{r}_j]^2 + 4 \tilde{z}_i \tilde{z}_j}}
    \biggr\}
    - \Xi (1+\Delta) \frac{Q}{q} \sum_{i,\alpha} \frac{1}{|\vmu{r}_i - \vmu{R}_{\alpha}|}
    + \sum_{i=1}^{N}  \frac{\Xi \, \Delta}{4 \tilde{z}_i}.
\end{multline}
The first sum corresponds to the interaction between pairs of counterions 
at positions ${\bf r}_i$
(taking into account the so-called ``image charges'' through the term 
proportional to  $\Delta$),
the second sum is the interaction between counterions and fixed surface charges
(the sum over $\alpha$ corresponds to all surface-ion
lattice vectors $\vmu{R}_{\alpha}$) and the third sum
is the interaction between counterions and their ``images.''
In our MC simulations we use Eq.~(\ref{energia})
for a typical number of $100$ counterions in conjunction with
periodic boundary conditions (implemented by
 the Lekner-Sperb technique\cite{lekner:91})
in order to minimize finite size effects\cite{moreira-netz:epl}; 
the results reported here are always taken from runs 
with $10^6$ Monte Carlo steps per particle.
In Fig.1b we show the effects of varying surface-charge modulation: For 
a ratio $D/a=0.24$ (to the right) the counterion snapshot shows a rather 
irregular configuration, with no or little visible correlation between 
surface charges (located beneath the nodes of the square lattice) and 
counter ions. For a ratio $D/a=0.06$ (to the left), 
on the other hand, the counterions
are strongly correlated with the square lattice of the surface ions 
(two counterions obviously have, driven by thermal fluctuations,
 escaped from their assigned lattice positions).
This is reflected 
by the normalized lateral counterion distribution 
$\tilde{\rho}_{yz}(\tilde{x})$, 
which for $D/a=0.24$ oscillates weakly around its mean value of unity, 
while for $D/a=0.06$ this distribution function is strongly peaked at
the positions of the surface ions.

\begin{figure}[t]
  \begin{center}
    \epsfig{file=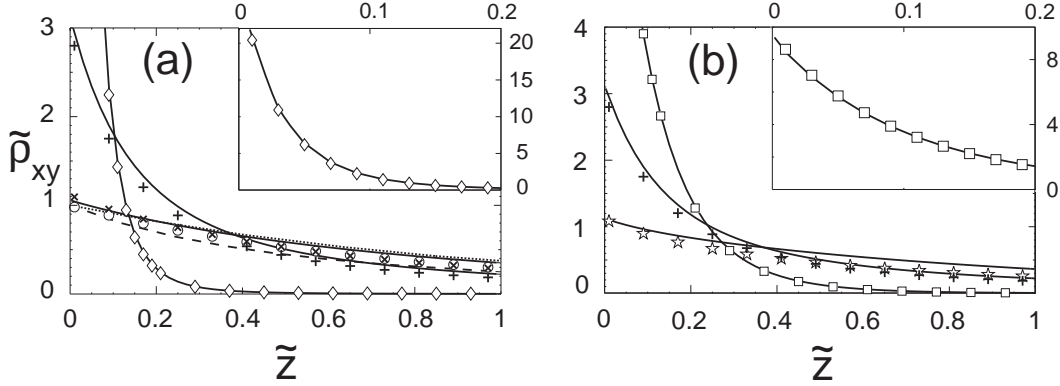, width=140mm}
  \end{center}
  \vspace{-8mm}
  \caption{Counterion density profile 
$\tilde{\rho}_{xy} $ 
(averaged in the $xy$--plane) as a function of $\tilde{z} \equiv z / \mu$
 for  $q=Q$ and without dielectric  jump ($\Delta=0$). a)
MC results (symbols) and SC predictions (full lines) for $\Xi=10$ and 
$D/a = 0.06$ (diamonds and inset), $D/a = 0.12$ (plus symbols), $D/a = 0.24$
(crosses).
The circles, dashed and dotted lines denote, 
respectively, the MC data, PB and SC predictions for $D/a = \infty$. 
b) MC results (symbols) and SC predictions (full lines) for
 $D/a = 0.12$ and $\Xi=1$ (stars), $\Xi=10$ (plus symbols) and $\Xi=100$ 
(squares and inset).  Error bars are smaller than symbol size.}    
  \label{fig:res_delta_0}
  \vspace{-5mm}
\end{figure}

Before we present the density profiles as a function of the distance from 
the substrate, we quickly recapitulate the SC theory. At 
leading order of an expansion in inverse powers of the coupling strength $\Xi$, 
which is equivalent to a virial expansion,
the normalized counterion density distribution is given by\cite{roland-sc}
\begin{equation}
  \label{SC}
  \tilde{\rho} ({\bf \tilde{r}}) = 
  \frac{\rho (\vmu{r})}{2 \pi \ell_B \sigma_s^2} = 
  \Lambda \, {\mathrm e}^{-\tilde{u}(\vmu{r})}+ {\cal O}(\Xi^{-1}).
\end{equation}
The factor $\Lambda$ is the fugacity of counterions,
determined by the normalization condition $\int {\mathrm d}\tilde{\bf r} \, 
\tilde{\rho} (\vmu{r}) /\tilde{L}^2  = 1$, where $\tilde{L}$ is the 
lateral system size,
equivalent to the condition of global electroneutrality\cite{roland-sc}. 
The function $\tilde{u}$, given by 
\begin{equation}
  \label{SC2}
  \tilde{u}(\vmu{r}) = -  \Xi (1+\Delta) \frac{Q}{q} 
  \sum_{\alpha} \frac{1}{|\vmu{r} - \vmu{R}_{\alpha}|}
    + \frac{\Xi \, \Delta}{4 \tilde{z}},    
\end{equation}
is the electrostatic potential created at position
$\vmu{r}$ by the surface ions at  lattice positions $\vmu{R}_{\alpha}$ 
(with $\tilde{Z}_{\alpha}=-\tilde{D}$) plus the
counterion--image interaction. 
In previous studies we demonstrated that the SC approach becomes 
quantitatively accurate for coupling strengths larger than $\Xi 
\approx 100$\cite{moreira-netz:epl}.
The discreteness of surface charges tends to decouple different
counterions from each other, as witnessed by the snapshots shown in 
Fig.1b. One would therefore expect corrections to the leading term of the systematic
expansion in Eq.(\ref{SC}), which come from correlations between 
counterions, to be weakened and 
the SC approach to perform even better in the presence of modulated 
surface charges, as indeed borne out by our data.
In the limit $D/a \rightarrow \infty$,
the smeared-out case,
 Eq.~(\ref{SC2}) reduces to $ \tilde{u}(\vmu{r}) =  (1+\Delta) \tilde{z}
 +\Xi \, \Delta /(4 \tilde{z})$.

We first discuss our results in the absence of a dielectric jump
($\Delta = 0$) and identical valences of surface and counterions ($Q=q$). 
In Fig.~\ref{fig:res_delta_0}a we fix $\Xi=10$   while
$D/a$ is varied; in Fig.~\ref{fig:res_delta_0}b we fix $D/a=0.12$ 
while $\Xi $ is varied. Clearly,
the counterion density profiles $\tilde{\rho}_{xy}(\tilde{z})$
 (which are averaged in the $xy$--plane)
are very sensitive to both the coupling constant $\Xi$ and the ratio $D/a$.
For $\Xi=10$ and $D/a=0.24$ (crosses in Fig.~\ref{fig:res_delta_0}a)
the discretization has a small effect, as the data almost coincide
 with the MC results  for the smeared-out case
  $D/a \rightarrow \infty$ (circles). 
With the rescaling of Fig.2a, 
the difference between the smeared-out SC and PB profiles 
(dotted and broken lines, respectively) is in fact rather small
compared to the effects of surface-charge modulation,
and the smeared-out data 
(circles) are somewhat in between the SC and PB predictions,
demonstrating that $\Xi=10$ is in the
crossover regime between strong and weak coupling\cite{moreira-netz:epl}.
  For $D/a = \infty$ and $\Delta =0$
  the contact-value theorem predicts a rescaled contact density of unity,
  $\tilde{\rho}_{xy}(\tilde{z}=0)=1$, as confirmed by simulations 
  at various values of $\Xi$\cite{moreira-netz:epl}.
However, as $D/a$  becomes smaller there is a greater accumulation of counterions
in the immediate vicinity of the charged surface, and the 
laterally averaged contact
density  can be several times larger than unity.
This is reflected by a high lateral correlation between 
counterions and the surface charges, see Fig.~\ref{fig:snapshots}b.
As $D/a$ becomes smaller, the SC predictions as defined in 
Eqs.~(\ref{SC}) and (\ref{SC2})  (solid lines 
in Fig.~\ref{fig:res_delta_0}a)
show progressively better agreement with the MC results. 
Similarly, in Fig.~\ref{fig:res_delta_0}b for fixed  $D/a=0.12$,
the contact density rises while 
the agreement between MC data and  SC predictions becomes quantitative
as $\Xi $ is increased.

\begin{figure}[t]
  \begin{center}
    \epsfig{file=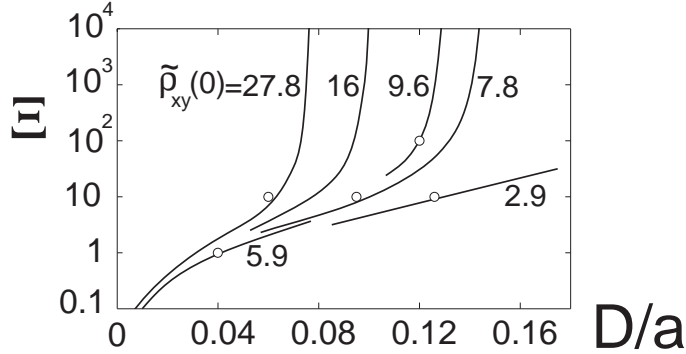, width=90mm}
  \end{center}
  \vspace{-8mm}
  \caption{Contour plots of constant contact counterion density 
$\tilde{\rho}_{xy}(\tilde{z}=0) $  as a function of the 
coupling constant $\Xi$ and the ratio $D/a$
for $\Delta=0$ and $Q=q$. The lines follow from the SC theory via 
Eq.~(\ref{SC}), while the open circles are obtained from MC data
via extrapolation leading to  $\tilde{\rho}_{xy}(0) = 2.9$, $5.9$, $7.8$, 
$9.6$ and $27.8$. For $D/a \rightarrow \infty $
as well as $\Xi \rightarrow 0$  the smeared-out assumption for the surface charge
distribution, characterized by $\tilde{\rho}_{xy}(0)=1$, is valid.}
  \vspace{-5mm}
  \label{fig:diagrama}
\end{figure}

These results are summarized in Fig.~\ref{fig:diagrama}, 
where contour plots of constant  contact density $\tilde{\rho}_{xy}(\tilde{z}=0) $ 
are shown as a function of $\Xi$ and $D/a$.
The lines were obtained from the SC theory via 
Eq.~(\ref{SC}) and agree well  with the 
MC results (open circles), obtained from density profiles
via extrapolation.
In the limit $D/a \rightarrow 0$ and finite coupling $\Xi$ 
the counterions collapse onto the surface charges and the density at 
contact diverges, while for fixed $D/a$ and in the limit 
$\Xi \rightarrow \infty$ the contact density saturates at a finite value.
 As $D/a$ grows (or $\Xi$  becomes
smaller), the smeared-out assumption becomes valid and $\tilde{\rho}_{xy}(0)$ 
approaches unity.

\begin{figure}[t]
  \begin{center}
    \epsfig{file=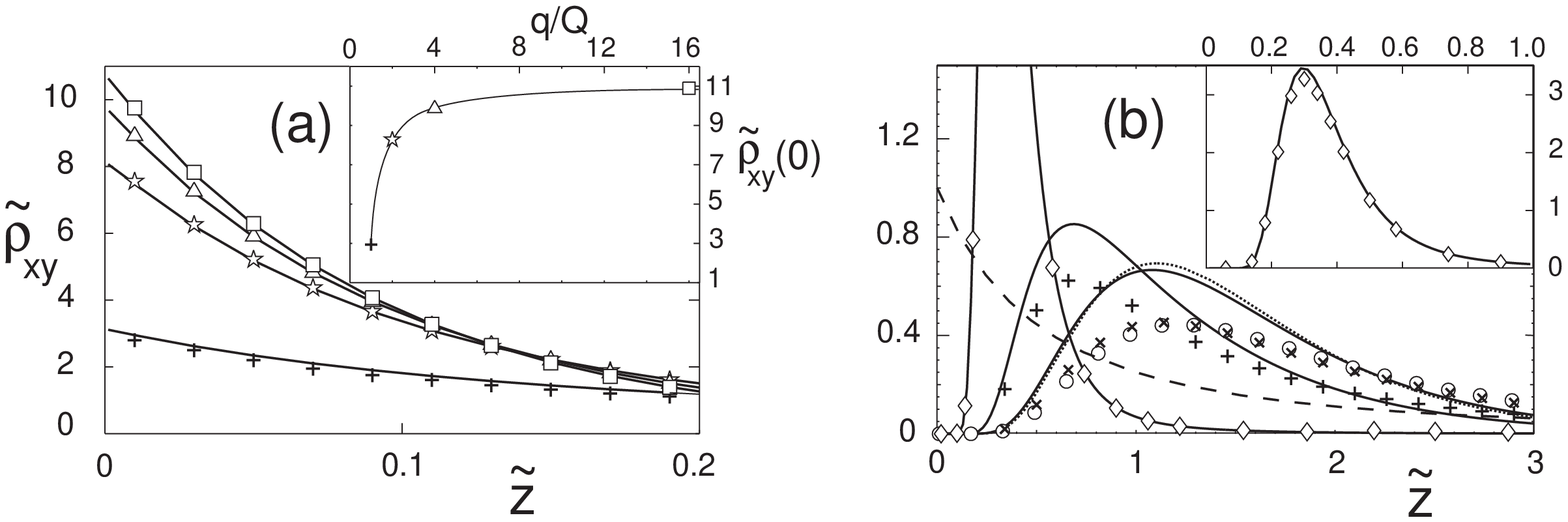, width=130mm}
  \end{center}
  \vspace{-8mm}
  \caption{a) Counterion density profiles $\tilde{\rho}_{xy}(\tilde{z})$
 for $\Xi Q^3/q^3=10$,
    $D/a = 0.12$, $\Delta=0$
 and different values of the valence ratio $q/Q$. 
    Full lines correspond to the SC predictions and symbols to
    MC results for $q/Q=1$ (plus symbols), $q/Q=2$ (stars), 
    $q/Q=4$ (triangles) and $q/Q=16$ (squares). 
    b) MC results (symbols) and SC predictions (full lines)
    for the counterion density profile for $\Xi=10$, 
    $q=Q$ in the presence of a dielectric jump, 
    $\Delta = 0.95$, for  $D/a = 0.06$ (diamonds and inset), 
    $D/a = 0.12$ (plus symbols) and $D/a = 0.24$ (crosses).
    The circles,  dashed and dotted lines denote, respectively, MC 
    data, PB and SC predictions in the smeared-out limit,  $D/a \rightarrow \infty$.}    
  \vspace{-5mm}
  \label{fig:res_2}
\end{figure}

We finally lift the restriction to symmetric surface/counterion valences 
($q=Q$) and to vanishing dielectric jump ($\Delta =0$).
In Fig.~\ref{fig:res_2}a we show density profiles for $\Xi Q^3 / q^3=10$, $D/a=0.12$,
$\Delta =0$ with varying valence ratios of $q/Q=1$ (plus symbols, already
featured in Fig.2), $q/Q=2$ (stars), $q/Q=4$
(triangles) and $q/Q=16$ (squares). The inhomogeneity of the surface charge distribution
becomes more important as the counterion valence increases. The SC results 
(solid lines) capture this trend and quantitatively agree with simulation data. 
The contact density $\tilde{\rho}_{xy}(0)$ saturates at a constant value
as $q/Q \rightarrow \infty$ (see inset).
For the case where a dielectric jump is present, we choose
$\varepsilon_> = 80$  and $\varepsilon_< = 2$, leading to $\Delta = 0.95$
and corresponding to a charged hydrocarbon substrate in contact with water.
Since $\Delta$ is positive, the interaction between the
counterions and their images is repulsive and competes with the
attractive interaction due to the surface charges, see \ Eq.~(\ref{energia}).
Consequently, the single-ion potential $\tilde{u}({\tilde{\bf r}})$, Eq.(\ref{SC2}), 
exhibits a   minimum at
a finite distance from the wall, and one expects the maximum in the density profile 
to be displaced from the wall, in agreement with previous 
results\cite{kjellander-marcelja:84,bratko-al:86} (however,
one should note that the resulting density profile depends strongly on details of the 
geometry and position of the dielectric boundary).
 Fig.~\ref{fig:res_2}b shows 
the counterion density profile for $\Xi=10$, $q=Q$, $\Delta = 0.95$ 
and the same values of $D/a$ as in 
Fig.~\ref{fig:res_delta_0}a, namely  $D/a = 0.06$, $0.12$ and $0.24$. 
The corresponding SC predictions (shown as solid lines) locate correctly the position
of the maximum in the density, though the agreement is not as good as 
without the dielectric jump for the larger values of $D/a$. 
The data for the smeared-out case,
$D/a =\infty$ (open circles) exhibit a maximum close to
$\tilde{z} = \sqrt{\Xi \Delta / (4 (1+\Delta))}$, the SC prediction, 
although the SC profile for this case (dotted line) 
overestimates the amount of counterions close to the wall. 
Obviously, the SC theory does not take into account the repulsion felt by a counterion 
due to the  presence of the images of other counterions, which explains 
why SC theory performs at the same value of $\Xi$ 
better in the absence of a dielectric jump  (compare Fig.2).
Higher order corrections to the asymptotic SC theory would therefore be 
needed to quantitatively match the data.
As is well-known, 
PB theory (shown as a broken line) is insensitive to the presence of a dielectric jump
in the smeared-out limit $D/a \rightarrow \infty $ and completely misses 
the shape of the density profile\cite{kjellander-marcelja:84,bratko-al:86}.

To conclude, we have studied both analytically and numerically
the distribution of counterions close to an inhomogeneously charged surface.
The charge inhomogeneity affects the counterion distribution strongly 
for pronounced charge modulation as well as when the coupling constant 
$\Xi$ is large.
As  summarized in Fig.~\ref{fig:diagrama} in the absence of a dielectric 
discontinuity ($\Delta =0$), the laterally averaged 
counterion contact density is much larger than 
the smeared-out value at decreasing $D/a$ (equivalent to a strongly 
charge-modulated substrate) and at
increasing coupling constant $\Xi$,
and counterions become strongly correlated with the
surface charges.
The same effect is reflected in the counterion density profiles, which 
under such conditions decay faster to zero as one moves away from the 
substrate, see Fig.2.

This has direct experimental consequences:
Very recently the counterion density profile (measured using ellipsometry)
at a surfactant monolayer with a high surface charge density 
(determined via  second-harmonic generation)
could be described by PB profiles only by assuming 
a fraction of counterions to be bound to the monolayer\cite{teppner},
similar to what we find. Our results suggest to reconsider
the traditional picture of a Stern layer, where the counterion surface 
concentration is 
subject to electrostatic and specific surface attraction and packing constraints, but where
the influence of lateral surface charge modulation 
(which is always present experimentally and leads to sizable effects as shown here) 
is typically neglected.
One should emphasize the experimental relevance of our parameters:
for a system with $\sigma_s \simeq 1/257$~{\AA}$^{-2}$ and divalent
counterions in water (which at room temperature leads to $\Xi=10$), 
the surface ions
are on average at a distance $a \simeq 16$~{\AA} from each other. 
For a minimal distance
between surface ions and counterions of $D \simeq 2$~{\AA} 
(noting that at close contact the hydrated water 
shell is stripped off) the ratio $D/a=0.12$ is obtained; 
as our results demonstrate,
the inhomogeneous character of the surface charges leads to pronounced 
deviations from the smeared-out case.

Other relevant factors, which we will consider in future work, include
 the possibility of the counterions to actually penetrate between the
surface ions, different surface charge patterns
(annealed and various quenched geometries), 
and additional interactions between surface groups and counterions
(like van der Waals or solvation forces) and between counterions themselves 
(most importantly hard core interactions, which to leading order
within the SC approach simply shift 
the density profile away from the surface).
The advantage of the currently employed model, though being quite crude and
neglecting a number of important effects, is that its simplicity allows
for a global analysis, encompassing different and limiting values
of the coupling strength and the degree of substrate charge modulation.
In a complementary study, similar numerical calculations have been recently 
analyzed within a modified PB approach\cite{Dima}.

\stars

We thank H.\ Motschmann and R.\ Teppner for useful discussions. AGM acknowledges
financial support from the DFG Schwerpunkt Polyelektrolyte.

\bibliographystyle{}

\end{document}